\documentclass[aps,preprint]{revtex4} 
\usepackage{graphics}
\begin{document}
\title{First-principles study of the dielectric and dynamical properties of orthorhombic CaMnO$_{3}$}

\author{Satadeep Bhattacharjee, Eric Bousquet and Philippe Ghosez}

\affiliation{Physique Th\'{e}orique des Mat\'{e}riqux, Universit\'e de Li\`{e}ge \\
All\'ee du 6 Ao\^ut 17, B-4000 Sart Tilman, Belgium} 
\begin{abstract}
The structural, dielectric and  dynamical properties of the low temperature 
antiferromagnetic orthorhombic phase of  CaMnO$_3$ have been computed 
from first principles 
using a density functional theory approach within the local spin density 
approximation. The theoretical structural parameters are in good agreement 
with experiment. The full set of  infrared and Raman zone-center phonons is 
reported and compared to experimental data.   It is shown that coherently with 
the anomalous Born effective charges and the presence of low frequency 
polar modes, the static dielectric constant is very large and highly anisotropic. 
\end{abstract}
\maketitle
\section{Introduction}

Magnetic perovskite oxides constitute an exciting subject of study. Indeed, 
the interplay between their structural, magnetic and transport properties  
make them fascinating from both experimental and theoretical point of views. 
For example, the mixed valent perovskite $Ca_{1-x}La_{x}MnO_{3}$ is one 
of the most studied material for its colossal magnetoresistance\cite{b0}. 
Also, the oxygen deficient manganites $LaMnO_{3-\delta}$ and $CaMnO_{3-\delta}$ 
are interesting for their transport and optical properties\cite{b01}. 

In this paper we report a first-principles study of  stoichiometric $CaMnO_{3}$. The ground-state crystal structure of $CaMnO_{3}$ is orthorhombic with space group Pnma\cite{b1}. The structure can be regarded as a distorted perovskite structure having four formula per unit\cite{b1,b2}. It is an insulator with an observed bandgap of about 3 eV\cite{b3}. The magnetic structure is antiferromagnetic and the $G-$type order is energetically the most favourable. The magnetic interactions between the Mn ions are due to superexchange interactions. Because of $Mn^{+4}$ configuration, there is no tendency to Jahn-Teller distortion. The  observed N\'eel temperature is about 130 K, which estimates the exchange energy to be 6.6 meV. 

Previous first-principles calculations on $CaMnO_{3}$ focused mainly on the electronic structure of the system. In this paper we extend the study and report results concerning the dielectric and dynamical properties of is ground-state orthorhombic phase. 

\section{Technical Details}

The first-principles simulations were performed according to the density functional theory scheme (DFT) within the local spin density approximation (LDA) and using the plane-wave implementation of the ABINIT package \cite{abinit}.
We used the Hartwigsen-Goedecker-Hutter \cite{hgh} (HGH) parametrization for the pseudopotentials, where the 3s and 3p orbitals were treated as valence for Mn and Ca atoms and 2s and 2p orbitals were considered as valence for O atoms.
The total number of valence electrons is therefore 15 for Mn, 10 for Ca and 6 for oxygen. 
Convergency was reached for an energy cutoff of 72 Hartree for the plane-wave expansion and a  $6\times4\times6$ $k$-point mesh for the Brillouin zone integration.
The phonon frequencies, Born effective charges and dielectric tensor were computed according to the density functional perturbation theory \cite{dfpt} (DFPT) scheme as implemented in the ABINIT package.

\section{Structural properties}

At high temperature, $CaMnO_3$ adopts a cubic perovskite structure.
At room temperature, it crystalizes in an orthorhombic \textit{Pnma} (N.62) 
structure with 20 atoms in the primitive unit cell and adopts a $G-$type 
antiferromagnetic configuration\cite{b1}. This orthorhombic phase results from the relaxation of three kinds 
of antiferrodistortive (AFD) instabilities which are related to rotations of oxygen 
octahedras around the Mn-O axis. In Glazer's notation, these tilts are described 
by a$^-$b$^+$c$^-$. Since these oxygen rotations do not strongly affect the high symmetric structure, the orthorhombic phase of CaMnO$_3$ can be 
described as a pseudocubic structure, the volume of which can be  estimated as $\sqrt2$a$_c \times$2a$_c \times \sqrt2$a$_c$ where a$_c$ is 
the pseudocubic cell parameter.

In our calculations, we imposed a $G-$type antiferromagnetic order and did the 
structural relaxation at fixed volume. The volume 
used corresponds a pseudocubic cell parameter of 3.73\AA\ which corresponds 
to the experimental cubic lattice constant \cite{b02} and also closely agree with the 
experimental pseudo-cubic lattice constant of the orthorhombic phase \cite{b1}. 
The cell shape and atomic positions were relaxed according to this constraint and 
the results are reported in TAB.\ref{struct}. 

The relaxed structure is in excellent agreement with the experimental data. The cell parameter $a$ and $b$ are only slightly overestimated while the $c$ parameter is slightly underestimated (errors of $\approx 0.5$ \%). The calculated atomic positions are also in good agreement with the experimental data, the distortion with respect to the ideal cubic positions being predicted with an accuracy similar to what is usually achieved in the class of $ABO_3$ compounds \cite{Rabe}.

From the inspection of the density of states, we found a theoretical electronic bandgap of 0.8 eV for our relaxed antiferromagnetic orthorhombic structure.  Although this strongly underestimates the  experimental value (3.1 eV \cite{gap}),  as usual within the LDA, our calculation properly reproduces the insulating character of the structure.  

\begin{table}[[htbp!]
\begin{tabular}{cccc}
\hline
\hline
 &\multicolumn{2}{c}{Orthorhombic}  & pseudocubic \\ 
 & This work & Exp.\cite{b1} & Exp. \cite{abrashev02} \\
\hline
a    & 5.287 & 5.279 & 5.275 \\
b    & 7.498 & 7.448 & 7.460 \\
c    & 5.235 & 5.264 & 5.275 \\
a$_c$& 3.730 & 3.726 & 3.730 \\
\hline
Ca (4c) & & & \\
x & 0.040 & 0.033 & 0.000 \\
y & 0.250 & 0.250 & 0.250 \\
z & -0.008 & -0.006 & 0.000 \\
Mn (4b) & & & \\
x & 0.000 & 0.000 & 0.000\\
y & 0.000 & 0.000 & 0.000\\
z & 0.500 & 0.500 & 0.500\\
O$_1$ (4c) & & &\\
x & 0.485 & 0.490& 0.500 \\
y & 0.250 & 0.250 & 0.250\\
z & 0.071 & 0.066 & 0.000\\
O$_2$ (8d) & & & \\
x & 0.287 & 0.287 & 0.250\\
y & 0.036 & 0.034 & 0.000\\
z & -0.288 & -0.288 & 0.250\\
\hline
\hline
\end{tabular}
\caption{Theoretical and experimental unit-cell parameters (\AA) and non-equivalent atomic positions (reduced coordinates) in the antiferromagnetic orthorhombic \textit{Pnma} structure of CaMnO$_3$. The Wyckoff position of the reported atoms are mentioned in brackets.}
\label{struct}
\end{table}

\section{Dielectric properties}

The Born effective charge tensors have been calculated for the the four 
atoms of TAB. \ref{struct}. The full tensors are as follows :

\begin{center}
\begin{tabular}{cccc}
$Z^*({Ca})=$ & $\left( \begin{tabular}{ccc} 
2.51  & 0.00 & 0.19 \\
0.00  &  2.43 & 0.00 \\
0.27 & 0.00 & 2.52 \\
\end{tabular}
\right)$
&
$Z^*({Mn})=$ & $\left( \begin{tabular}{ccc} 
6.82  & -0.37 & -0.72 \\
0.07  &  5.78 & 1.42 \\
-0.69 & -1.54 & 6.56 \\
\end{tabular}
\right)$
\\
&$\left[ 2.29\ \ 2.74\ \ 2.43 \right]$ &&$\left[ 7.41\ \ 6.05\ \ 5.70 \right]$  \\
&\\
$Z^*({O_1})=$& $\left( \begin{tabular}{ccc} 
-1.73  & 0.00 & 0.06 \\
0.00  &  -5.11 & 0.00 \\
-0.20 & 0.00 & -1.76 \\
\end{tabular}
\right)$
&
$Z^*({O_2})=$& $\left( \begin{tabular}{ccc} 
-3.80  & -0.08 & -1.94 \\
-0.11  &  -1.55 & -0.01 \\
-1.97 & 0.00 & -3.61 \\
\end{tabular}
\right)$
\\
& $\left[ -1.67\ \ -1.82\ \ -5.11 \right]$ &&$\left[ -5.66\ \ -1.77\ \ -1.53 \right] $\\
\end{tabular}
\end{center}

Below each tensor, the main values of the symmetric part of $Z^*$ is also mentioned,
for direct comparison with (i) the nominal atomic charges ($Z_N(Ca) = +2$, $Z_N(Mn) = +4$ and $Z_N(O) = -2$) and (ii)  the Born effective charges in the cubic phase at $a_c = 3.73$\AA ($Z^*_C(Ca) = +2.61$, $Z^*_C(Mn) = +7.43$, $Z^*_C(O_{\perp}) = -2.55$ and $Z^*_C(O_{\parallel}) = -4.94$). The trends are rather similar to what was previously reported for the isostructral $CaTiO_3$ \cite{orthoCTO}. As usual within this class of ABO$_3$ compounds \cite{Ghosez-Z,Rabe}, we notice the anomalously large values of the Mn and O charges. As usual also,  in comparison to the highly symmetric cubic structure, there is a  reduction of the anomalous values in the low symmetry phase,  at the exception of $Z^*_{yy}(O_1)$. Finally, we notice the existence of small asymmetric contributions, as allowed by symmetry in the orthorhombic phase.

Although its value is not expected to be very accurately predicted within the LDA \cite{Ghosez-eps}, we also report the calculated optical dielectric tensor :

\begin{center}
\begin{tabular}{cc}
$\epsilon_{\infty} = $ & $\left( \begin{tabular}{ccc} 
11.3  & 0.0 & 0.0 \\
0.0  &  13.1 & 0.0 \\
0.0 & 0.0 & 10.8 \\
\end{tabular}
\right)$
\end{tabular}
\end{center}

The calculated tensor is only slightly anisotropic. Its elements are significantly larger than in related compounds like $CaTiO_3$ ($\epsilon_{\infty} \approx 6$ \cite{orthoCTO}), coherently with the smaller LDA bandgap of $CaMnO_3$. The optical dielectric tensor in the orthorhombic phase does not differ significantly from that previously reported in the ideal cubic structure ($\epsilon_{\infty} =10.43$ from our calculations and $\epsilon_{\infty} = 11.25$ from Ref. \onlinecite{filippetti02}).  We notice however that it is slightly larger than in the cubic phase, which is a trend  different from that reported for orthorhombic $CaTiO_3$ in which  $\varepsilon_{\infty}$ decreases 
when non-polar distortions are frozen into the structure\cite{orthoCTO}.

\section{Dynamical properties}

The irreducible representation in the orthorhombic \textit{Pnma} phase CaMnO$_3$ at the $\Gamma$ point is:  

$$7A_{g}\oplus5B_{1g}\oplus7B_{2g}\oplus5B_{3g}\oplus10B_{1u}\oplus8B_{2u}\oplus10B_{3u}\oplus8A_{u}$$.

Over this decomposition, 3 modes are acoustic (symmetries $B_{1u}$, $B_{2u}$, $B_{3u}$), 8 are silent (symmetry $A_u$), 24 are Raman (R) active (symmetries $A_{g}$, $B_{1g}$, $B_{2g}$ and $B_{3g}$) and the last 25 modes are infrared (IR) active (symmetries $B_{1u}$, $B_{2u}$ and $B_{3u}$).
According to the structure defined in TAB.\ref{struct}, the IR $B_{3u}$ modes are polarized along the \textit{x} direction, $B_{2u}$ along \textit{y} direction and $B_{1u}$ along \textit{z} direction.
The eight $A_u$ silent modes, not further discussed below, are calculated at the frequencies of 123, 140, 179, 220, 313, 392, 438 and 466 cm$^{-1}$

\subsection{Raman active modes}

We report in TAB.\ref{raman} the calculated frequencies of the Raman active modes and  a related new assignment of the experimental data. The latter is compared to the assignment previously proposed from shell-model results in Ref.\cite{abrashev02}.

Experimentally, the main observed modes are those of  A$_g$ symmetry.
10 A$_g$ modes were measured at 150, 160, 184, 243, 278, 322, 382, 438, 487 and 615 cm$^{-1}$ while only 7 modes are expected from group theory.
The frequency at 615 cm$^{-1}$ was unambigously assigned to impurity and the modes at 382 and 438 cm$^{-1}$ were kept aside since their frequencies differ strongly from those of the shell model.  Below 200 cm$^{-1}$ the assignment was particularly ambiguous since two modes were calculated at 154 and 200 cm$^{-1}$, while three were oberved in the spectra.

From our calculation, the experimental lines at 243, 278 and 322 cm$^{-1}$ can be assigned to the A$_g$ modes at 250, 275 and 314 cm$^{-1}$  respectively. For the low frequencies, the assignment can be significantly improve if we assign to the calculated modes at 152 and 167 cm$^{-1}$, the experimental modes at 150 and 160 cm$^{-1}$  rather than those at 160 and 184 cm$^{-1}$ as previously proposed (i.e. rulling out the experimental mode at 184 cm$^{-1}$ instead of the one at 150 cm$^{-1}$). For the high frequency A$_g$ modes, the overall agreement between experimental and theoretical data is the best if we assign the experimental line at 438 cm$^{-1}$ (not  assigned in Ref.\cite{abrashev02})  to the calculated mode at 450 cm$^{-1}$ and the one at  487 cm$^{-1}$ to that calculated mode at 504 cm$^{-1}$.

The experimental line at 179 cm$^{-1}$, assigned to a B$_{1g}$ mode in Ref.\cite{abrashev02}, is in good agreement with our calculation (189 cm$^{-1}$). Amongst the two experimental lines at 320 and 564 cm$^{-1}$ previously assigned to of B$_{3g}$ modes \cite{abrashev02}, the first one is reproduced within our calculation with a very good accuracy  (320 cm$^{-1}$), contrary to the second for which we get a theoretical frequency of 469 cm$^{-1}$). Since the assignment between B$_{1g}$ and B$_{3g}$ modes is rather ambiguous, we propose that the experimental line at 564 cm$^{-1}$ might in fact correspond to the  B$_{1g}$ mode calculated at 595 cm$^{-1}$.

Finally, from Ref.\cite{abrashev02}, it seems that the B$_{2g}$ modes are theoretically predicted with less accuracy since modes calculated with the shell model at 180 and 425 cm$^{-1}$ are associated to the lines measured at 258 and 465 cm$^{-1}$. From our first-principles calculation, this agreement seems significantly better and we propose to assign the experimental line at 258  cm$^{-1}$ to the theoretical mode at 227  cm$^{-1}$ and the line at 465  cm$^{-1}$ to the theoretical mode at  465 cm$^{-1}$.

\begin{table}[[htbp!]
\begin{tabular}{ccccccc}
\hline
\hline
Symmetry & \multicolumn{2}{c}{Present} && &\multicolumn{2}{c}{Reference \cite{abrashev02}} \\
 & FP     & Exp. \cite{abrashev02}           &&        & SM                 & Exp.\\
\hline
$A_{g}$ & 152 & 150 &&& 154 & 160  \\
$A_{g}$ & 167 & 160 &&& 200 & 184  \\
$B_{2g}$& 172 & -   &&& 148 & -    \\
$B_{2g}$& 180 & -   &&& 232 & 258  \\
$B_{1g}$& 189 & 179 &&& 178 & 179  \\
$B_{3g}$& 203 & -   &&& 290 & -    \\
$B_{2g}$& 227 & 258 &&& 292 &      \\
$B_{1g}$& 241 & -   &&& 281 & -    \\
$A_{g}$ & 250 & 243 &&& 242 & 243  \\
$A_{g}$ & 275 & 278 &&& 299 & 278  \\
$A_{g}$ & 314 & 322 &&& 345 & 322  \\
$B_{3g}$& 320 & 320 &&& 304 & 320  \\
$B_{1g}$& 330 & -   &&& 354 &      \\
$B_{2g}$& 372 & -   &&& 366 & -    \\
$B_{2g}$& 425 & -   &&& 453 & 465  \\
$B_{3g}$& 434 & -   &&& 459 & -    \\
$A_{g}$ & 450 & 438 &&& 467 & 487  \\
$B_{2g}$& 465 & 465 &&& 485 &      \\
$B_{3g}$& 469 & -   &&& 541 & 564  \\
$B_{1g}$& 488 & -   &&& 536 & -    \\
$A_{g}$ & 504 & 487 &&& 555 & -    \\
$B_{1g}$& 595 & 564 &&& 743 & -    \\
$B_{2g}$& 655 & -   &&& 749 & -    \\
$B_{3g}$& 674 & -   &&& 754 & -    \\
\hline
\hline
\end{tabular}
\caption{Comparison of the calculated and experimental frequencies (cm$^{-1}$) of the Raman modes of the antiferromagnetic orthorhombic phase of $CaMnO_3$.
The first column labels the symmetry of each mode.  The second and third  columns correspond to the present first-principles (FP) calculations and the related assignment of the experimental data of Ref. \cite{abrashev02}.  The fourth and fifth columns corresponds to the shell-model (SM) calculations and related assignment in Ref.\cite{abrashev02}.}
\label{raman}
\end{table}

\subsection{IR active modes}

Table \ref{IR} summarizes the calculated IR frequencies of the TO modes.
Unfortunately no IR measurement on monocrystal in the  orthorhombic 
phase of CaMnO$_3$ was found to compare with our results.
An IR spectra was reported in Ref.\cite{fedorov99} but on a polycrystalline 
sample and no symmetry attribution was reported for the 15 observed 
frequencies, which makes  comparison very difficult considering the high 
number of modes. The only remark that can be done concern the high frequency
part of the spectrum. The three highest frequencies are measured in 
Ref.\cite{fedorov99} at 533, 580 and  628 cm$^{-1}$, which deviate strongly
 with our results where the maximum frequency of IR modes is calculated at 
 504 cm$^{-1}$. This could eventually be related to inaccuracies in our 
 calculations but our calculation also suggest that these experimental frequencies 
 might also likely correspond to combination of modes  (for exemple, the highest measured frequency 628 cm$^{-1}$ can  be recover as being the exact sum of the modes calculated  at 305 cm$^{-1}$ and 323 cm$^{-1}$).

\begin{table}[[htbp!]
\begin{tabular}{ccccc}
\hline
\hline
Symmetry & $\omega$ & $\overline{Z}^*_{m,\alpha\alpha}$ &$S_{m}^{\alpha\alpha}$& $\varepsilon_{0,m}^{\alpha\alpha}$\\
                &[cm$^{-1}$]  & [$|e^-|$] & [10$^{-4} a.u.$] &[--] \\
\hline
$B_{3u}$& 101 & 15.95  & 56.62  & 238.41\\
$B_{1u}$& 147 & 10.74  & 17.68  & 35.09\\
$B_{1u}$& 150 & 13.63  & 34.13  & 65.37\\
$B_{3u}$& 153 & 3.97   & 2.22  & 4.06\\
$B_{2u}$& 155 & 2.03   & 0.43  & 0.77\\
$B_{2u}$& 179 & 5.60   & 8.50  & 11.40\\
$B_{2u}$& 208 & 14.26  & 37.69  & 37.67 \\
$B_{1u}$& 215 & 5.6    & 6.27  & 5.86\\
$B_{3u}$& 216 & 3.10   & 1.70  & 1.57\\
$B_{3u}$& 234 & 5.06   & 5.02  & 3.94\\
$B_{2u}$& 251 & 4.60   & 5.53  & 3.80\\
$B_{3u}$& 287 & 3.4    & 3.53  & 1.85\\
$B_{1u}$& 290 & 0.59   & 0.09  & 0.04\\
$B_{2u}$& 324 & 2.70   & 0.98  & 0.37\\
$B_{1u}$& 326 & 2.59   & 1.92  & 0.78\\
$B_{3u}$& 340 & 1.40   & 0.36  & 0.13\\
$B_{1u}$& 345 & 2.36   & 1.43  & 0.51\\
$B_{1u}$& 381 & 6.18   & 6.13  & 1.80\\
$B_{1u}$& 414 & 0.06   & 0.00  & 0.00\\
$B_{3u}$& 423 & 1.92   & 1.07  & 0.25\\
$B_{2u}$& 437 & 1.89   & 1.19  & 0.26\\
$B_{3u}$& 469 & 4.22   & 5.48  & 1.07\\
$B_{2u}$& 488 & 4.14   & 5.75  & 1.04\\
$B_{3u}$& 495 & 1.4    & 0.54  & 0.09\\
$B_{1u}$& 504 & 3.4    & 3.84  & 0.65\\
\hline
\hline
\end{tabular}
\caption{Symmetry and frequency of the 25 IR modes. The mode 
effective charges ($\overline{Z}^*_{m,\alpha\alpha}$, as defined in  
Ref. \cite{Lee}) and the oscillator  strengths  ($S_{m}^{\alpha\alpha}$) 
are also provided for each mode as well as the contribution to the static 
dielectric constant $\varepsilon_{0,m}^{\alpha\alpha}$.  From our
 conventions, $ \alpha\alpha= xx$ for $B_{3u}$ modes, $\alpha\alpha = yy$ 
 for $B_{2u}$ modes and $\alpha\alpha = zz$ for $B_{1u}$ modes.}
\label{IR}
\end{table}

In addition to the frequencies, we also report in TAB.\ref{IR} the mode effective charges (\cite{Lee}), the oscillator strengths and the contribution of each polar mode to the static dielectric tensor. The total static dielectric tensor can be decomposed as follows:
\begin{eqnarray}
\epsilon_0^{\alpha\beta}=\epsilon_{\infty}^{\alpha\beta}+\sum_{m}\epsilon_{0,m}^{\alpha\beta}
\end{eqnarray}
where $\alpha$ and $\beta$ are the cartesian directions (x, y or z), $\epsilon_{\infty}$ is the optical dielectric tensor and $\epsilon_{0,m}$ is the contribution to the dielectric constant of each individual phonon mode $m$. This latter contribution  is computed from the following relation:
\begin{eqnarray}
 \epsilon_{0,m}^{\alpha\beta}={4\pi\over\Omega}{S_m^{\alpha\beta}\over \omega_m^{2}}
\end{eqnarray}

where $\Omega$ is the volume of the cell, $S_m^{\alpha\beta}$ and $\omega_m$ are respectively the oscillator strength and the frequency of the mode $m$.

The computed static dielectric tensor is reported below :

\begin{center}
\begin{tabular}{cc}
$\epsilon_{0} = $ & $\left( \begin{tabular}{ccc} 
262  & 0.0 & 0.0 \\
0.0  & 68 & 0.0 \\
0.0 & 0.0 & 120 \\
\end{tabular}
\right)$
\end{tabular}
\end{center}

Not only the static dielectric tensor takes values significantly larger than the optical dielectric tensor but, contrary to the latter, it is also highly anisotropic. Some insight into this result is provided from the inspection of TAB \ref{IR}. Along the $x$ direction, the static dielectric constant ($\epsilon_{0}^{xx} = 262$)  appears as mainly due to the contribution of  the lowest B$_{3u}$ mode at 101cm$^{-1}$ (238), which combine giant mode effective charge and oscillator strength and a low frequency. Along the $z$ direction, there are still highly polar mode (B$_{1u}$) with giant mode effective charges 
but at higher frequencies so that the static dielectric constant is smaller. Along the $y$ direction, the most polar mode (B$_{2u}$) is still at larger frequency.  

\section{Conclusion}

The structural, dielectric and dynamical properties of the anti-ferromagnetic orthorhombic phase of $CaMnO_3$ have been studied from first-principles. The relaxed structure is in good agreement with experimental data. As within the cubic phase, the Born effective charges of Mn and O are highly anomalous. The whole set of zone-center phonon modes has been computed and a new assignment of experimental data as been proposed. The static dielectric tensor has also been obtained : it shows amplitudes comparable to CaTiO$_3$ and is highly anisotropic. Both the inspection of the phonon modes and static dielectric tensor emphasize that, although from the structural point of view the orthorhombic phase can be considered as a pseudo-cubic structure, from the dielectric and dynamical point of view, it is highly anisotropic.

\section{Acknowledgement}
This work was supported by the European STREP MaCoMuFi, the VolkswagenStiftung and the European Network of Excellence FAME-EMMI.

\end{document}